\documentclass[aps,pra,showpacs,twocolumn,amsmath,amssymb]{revtex4}
\usepackage{graphicx,bbm}

\newcommand{\xbra}[1]{{}_\sharp{\langle #1 \vert}}
\newcommand{\ket}[1]{{\vert #1 \rangle}}
\newcommand{\xket}[1]{{\vert #1 \rangle}_\sharp}
\newcommand{\braket}[2]{\langle #1 \vert #2 \rangle}
\newcommand{\xbraket}[2]{{}_\sharp\langle #1 \vert #2 \rangle_\sharp}
\newcommand{\xave}[1]{{\langle #1 \rangle_\sharp}}

\newcommand{\ee}{ {\rm e} }
\newcommand{\ii}{ {\rm i} }
\newcommand{\dd}{ {\rm d} }
\newcommand{\ZZ}{\mathbb{Z}}
\newcommand{\RR}{\mathbb{R}}
\newcommand{\CC}{\mathbb{C}}
\newcommand{\cc}{ {\hat c} }
\newcommand{\nn}{ {\hat n} }
\newcommand{\ww}{ {\hat w} }

\newcommand{\ve}[1]{{\mathbf #1}}
\def\tr{{{\rm tr}}}
\def\one{\mathbbm{1}}
\def\AA{{\frak A}}
\def\one{\mathbbm{1}}

\begin{document}

\title{Operator space entanglement entropy in transverse Ising chain}
\author{Toma\v{z} Prosen}
\author{Iztok Pi\v{z}orn}
\affiliation{Department of physics, FMF, University of Ljubljana,
Jadranska 19, SI-1000 Ljubljana, Slovenia}

\begin{abstract}
	The efficiency of time dependent density matrix renormalization group methods is intrinsically 
	connected with the rate of entanglement growth.
        We introduce a new measure of entanglement in the space of
	operators and show, for transverse Ising spin $1/2$ chain, that
	the simulation of observables, contrary to simulation of
	typical pure quantum states, {\em is} efficient for initial local operators.
        For initial operators with a finite index in Majorana representation, the
	operator space entanglement entropy saturates with time to a level which is
	calculated analytically, while for initial operators with infinite index the growth
	of operator space entanglement entropy is shown to be logarithmic.
\end{abstract}


\maketitle

\section{Introduction}
\label{sect:introduction}

The entanglement is an intrinsic property of composite quantum systems
and represents a cornerstone in quantum information theory \cite{nielsenbook}.
It is important to understand the role of quantum entanglement in classical
manipulation of quantum objects, and to quantify the degree of entanglement.
Although the question of quantification is not clearly resolved, the 
quantum information theory offers several measures \cite{eisertJMO, plenio0504163} of entanglement.
Quantum information theory also gave a new birth or fresh interpretation of 
a class of methods for
numerical simulation of many-body quantum systems which, due to the
exponential growth of Hilbert space, cannot be manipulated using exact diagonalization.
The methods originally known as density matrix renormalization group 
(DMRG) \cite{whitePRL69} deploy the fact that 
many degrees of freedom are redundant in quantum state description;
the system is therefore adequately described by taking into account
maximally entangled components only.
Thus, sufficiently slow growth of the entanglement is of crucial importance.
DMRG enjoyed remarkable success in determining the ground state properties 
of large one-dimensional quantum models, for which the
degree of entanglement scales at most logarithmically 
with size \cite{latorre,jinJSP116,keatingCMP252,holzhey,osborne,osterloh},
however its {\em time-dependent} version (tDMRG) \cite{vidalPRL91, whitePRL93}
is often plagued by abundance of entanglement with time evolution \cite{prosenPRE75}. 
For efficient classical simulation of many body quantum dynamics using tDMRG it is required that 
the computational costs grow polynomially in time meaning that the degree of entanglement of
any quantum object which can be represented as an element of a scalable tensor product 
Hilbert space (either a pure state, or a mixed state/operator, etc)
must grow no faster than logarithmically. 
It was recently shown that this is generically not the
case for {\em quantum chaotic} Ising spin chain in a tilted magnetic field where 
the entanglement entropies grow linearly and hence the computation costs 
increase exponentially in time \cite{prosenPRE75}. 

In this paper we shall consider the model of quantum Ising chain
in transverse magnetic field which is integrable and 
an explicit analytical solution exists. 
Calabrese and Cardy \cite{calabreseJSM4} have shown numerically that
the growth of entanglement entropy is {\em linear} for evolution of pure initial states  
which are ground states of quenched hamiltonians; see also Ref.\cite{chiaraJSM6}.
However, from efficiency of tDMRG for time evolution 
of local operators \cite{prosenPRE75} one may conclude that entanglement entropy
computed in the space of operators grows only logarithmically.
Here we address this problem theoretically using the idea 
\cite{prosenPRE60} of re-formulating the Heisenberg evolution in an algebra of operators
in terms of a Schr\" odinger evolution generated by a different - adjoint hamiltonian
acting on the Hilbert space of operator algebra. We show that
operator space entanglement entropy saturates in time for initial local operators of
finite index (precise definitions follow) and explicitly compute the saturation values in the
critical case. Further, for initial local operators of infinite index we give accurate numerical evidence and 
a theoretical hint that the growth is logarithmic (in thermodynamic limit) with prefactor 
$1/6$ in critical, or $1/3$ in non-critical case.

We note that, to best of our knowledge, the entanglement in operator space is a new concept
which has not yet been considered theoretically, and it is clearly not equivalent to the 
conventional concept of entanglement of density operators as discussed in sect.\ref{sect:conclusions}.
Yet, it is the one which we expect to be more closely related to computational complexity of time-evolution in infinite interacting quantum systems.

\section{Fermion representation of dynamics in operator space}

The dynamics of a transverse Ising chain of length $2L$ is described  
in terms of canonical Pauli matrices $\sigma^\alpha_j$ for sites 
$j\in \ZZ_{2L}\equiv\{ -L+1,\ldots,0,1,\ldots,L\}$ and the hamiltonian
\begin{equation}
	H = \sum_{j=-L+1}^{L-1} \sigma_j^{\rm z} \sigma_{j+1}^{\rm z} + 
 	h \sum_{j=-L+1}^{L} \sigma_j^{\rm x}
	\label{eq:H}
\end{equation}
with open boundary conditions, which
 can be diagonalized by means of
Jordan-Wigner transformation and introduction of 
Majorana fermion operators \cite{brandtjacoby, prosenPRE60},
	$X_n = \big(\prod_{j<n}\sigma_j^{\rm z}\big) \sigma_n^{\rm x}$ and 
	$Y_n = \big(\prod_{j<n}\sigma_j^{\rm z}\big) \sigma_n^{\rm y}$
fulfilling the anti-commutation relations
$\{X_i,X_j\}=\{Y_i,Y_j\}=2\delta_{ij}$ and $\{X_i,Y_j\}=0$.
Heisenberg equations of motion $\dd A/\dd t = \ii [H,A]$ for
 Majorana operators can be written:
\begin{equation}
	{\dot X_n} = 
	2(Y_{n-1}\!-\! h Y_n ),
	\qquad
	{\dot Y_n} = -2(X_{n+1}\!-\! h X_n).
	\label{eq:dXndt}
\end{equation}
An operator corresponding to an arbitrary physical observable can be
written as a superposition 
of products of Majorana 
operators $X_j$, $Y_j$, namely
$P_{\mathbf{n},\mathbf{n'}} = X_{-L+1}^{n_{-L+1}} Y_{-L+1}^{n'_{-L+1}}\cdots %
X_L^{n_{L}} Y_L^{n'_{L}}$ 
with powers $n_j,n'_j \in \{0,1\}$. 
A set of $4^{2L}$ operators $\{ \xket{P_{\mathbf{n},\mathbf{n'}}} \}$
spans an orthonormal basis of a Hilbert space, namely the matrix
alebra $\AA=\CC^{2^{2L}\times 2^{2L}}$,
with an inner product $\xbraket{A}{B}= 2^{-2L} \tr A^\dagger B$.
$\AA$ can be conveniently interpreted as a Fock space of $2L$ 
{\em adjoint fermions} (we shall call them a-fermions) with pseudo-spin 
(distinguishing between Majorana $X_j$ and $Y_j$ operator).
An arbitrary operator $A$ is then a-fermion state
$\xket{A} = \sum_{\mathbf{n},\mathbf{n'}} a_{\mathbf{n},\mathbf{n'}}
\xket{P_{\mathbf{n},\mathbf{n'}}}$. A-fermi operators over
$\AA$, $\cc_{j\uparrow},\cc_{j\downarrow}$, can be introduced by
\begin{eqnarray}
\cc_{j\uparrow} \xket{P_{\mathbf{n},\mathbf{n'}}} &=& 
n_j \xket{X_j P_{\mathbf{n},\mathbf{n'}}}  \\
\cc_{j\downarrow} \xket{P_{\mathbf{n},\mathbf{n'}}} &=& 
n'_j \xket{Y_j P_{\mathbf{n},\mathbf{n'}}}  
\end{eqnarray}
satisfying canonical anti-commutation relations.

The index of an operator $P_{\mathbf{n},\mathbf{n'}}$ is defined as 
$I_{\mathbf{n},\mathbf{n'}} = \sum_{j}(n_j+n'_j)$ 
and for index-one operators eq. (\ref{eq:dXndt}) rewrites: 
\begin{eqnarray}
	\frac{\dd}{\dd t} \xket{X_n} &=&  
	2(\cc_{n-1\downarrow}^\dagger - h \cc_{n\downarrow}^\dagger) 
	\cc_{n\uparrow} \xket{X_n}\\
	\frac{\dd}{\dd t} \xket{Y_n} &=&  
	-2(\cc_{n+1\uparrow}^\dagger - h \cc_{n\uparrow}^\dagger) 
	\cc_{n\downarrow} \xket{Y_n}
	\label{eq:dXndYn2}
\end{eqnarray}
which can be interpreted as a Schr\" odinger equation
$(\dd/\dd t) \xket{A} =
-\ii \hat{\mathcal{H}}\xket{A}$
 for the adjoint hamiltonian
\begin{equation}
	\hat{\mathcal{H}} = 
	 2 {\rm i}\!\!\sum_{n\in\ZZ_{2L}}\!\!\!\Big[ 
	(\cc_{n-1\downarrow}^\dagger \cc_{n\uparrow} - \cc_{n+1\uparrow}^\dagger \cc_{n\downarrow})
	+ h ( \cc_{n\uparrow}^\dagger \cc_{n\downarrow} - \cc_{n\downarrow}^\dagger c_{n\uparrow})
\Big].
\label{eq:hamiltonianHcal}
\end{equation}
Since the adjoint time-evolution is a {\em homomorphism} the 
Schr\" odinger equation extends to arbitrary element of the operator algebra $\xket{A}\in\AA$.
Note that the number of a-fermions, $\hat{\mathcal{N}}=\sum_{n,s} \cc^\dagger_{n,s} \cc_{n,s}$ is conserved, unlike the number of ordinary Majorana fermions.

\section{Operator space entanglement entropy}

It is clear that classical simulability of quantum states is quantified by the entanglement
entropy of half-half (or worst case) bipartition of the lattice. However, for simulability of
quantum observables (or density operators of mixed states), the decisive quantity is an
analog of entanglement entropy defined for an arbitrary element of operator algebra
$\AA \ni \xket{A} = \sum_{\mathbf{n},\mathbf{n'}} a_{\mathbf{n},\mathbf{n'}}
\xket{P_{\mathbf{n},\mathbf{n'}}}$, with the adjoint reduced density matrix
\begin{eqnarray}
&& R_{(n_{-\!L\!+\!1},n'_{-\!L\!+\!1},\ldots n_0,n'_0),(m_{-\!L\!+\!1},m'_{-\!L\!+\!1},\ldots m_0,m'_0)} =  \nonumber \\
&& \sum_{n_1,n'_1,\ldots n_L,n'_L} a_{(n_{-\!L\!+\!1},\ldots n_0,n_1,\ldots n_L),(n'_{-\!L\!+\!1},\ldots n'_0,n'_1,\ldots n'_L)} \times \nonumber \\
&& a_{(m_{-\!L\!+\!1},\ldots m_0,n_1,\ldots n_L),(m'_{-\!L\!+\!1},\ldots m'_0,n'_1,\ldots n'_L)}^*,
\end{eqnarray}
namely 
\begin{equation}
S = -\tr R \ln R.
\label{eq:SreducedR}
\end{equation}
For a spin 1/2 chain it is perhaps more natural to use a set of $4^{2L}$ Pauli operators
$Q_{s_{-\!L\!+\!1},\ldots s_L} = \sigma_{-\!L\!+\!1}^{s_{-\!L\!+\!1}}\cdots \sigma^{s_L}_L$, where
$s_j\in\{0,{\rm x},{\rm y},{\rm z}\}$, $\sigma^0\equiv \one$, as a physical basis of 
operator algebra $\AA$, and define bi-partition and entanglement entropy with 
respect to $Q_{\ve{s}}$. However, it is easy to show that the result is identical to 
(\ref{eq:SreducedR}) 
since the transformation between the bases
$\{ P_{\ve{n},\ve{n}'}\}$ and $\{ Q_{\ve{s}} \}$ is a simple {\em permutation} of multiindices
$(\ve{n},\ve{n}')\leftrightarrow \ve{s}$ (with multiplications by $\pm 1$), and even though it is 
{\em non-local} it maps first $L$ a-fermions to only first $L$ spins and vice versa.
  
Let us now try to compute {\em time-dependence} of operator space entanglement entropy
$S(t)$ for some simple initial operators $A$.
For convenience, we introduce staggered a-fermi operators $\ww_j$, 
$j\in \ZZ_{4L}=\{ -2L+1,\ldots,0,1,\ldots,2L\}$, such that
$\ww_{2n-1} \equiv \cc_{n\uparrow}$ and $\ww_{2n} \equiv \cc_{n\downarrow}$. 
Any operator acting solely in a space of first $L$ a-fermions (or first $L$ spins)
 can be expressed in terms of $2L$ anti-comuting operators $\ww_j$ with 
 $j \in \ZZ_{2L}^- \equiv \{-2L+1, \ldots, -1, 0\}$.
We follow Ref.\cite{jinJSP116} and express $2^{2L}$ eigenvalues of adjoint reduced density matrix
$R$, as $\rho_{\ve{n}} = \prod_j\left( n_j \gamma_j + (1\!-\!n_j)(1\!-\!\gamma_j)\right)$, $n_j\in\{0,1\}$
where $\gamma_j$ are eigenvalues of time-dependent $2L\times 2L$ correlation matrix
\begin{equation}
	\Gamma_{mn}(t) = \xbra{A} \ww_m^\dagger(t) \ww_n(t) \xket{A},
	\quad m,n \in \ZZ_{2L}^-
\end{equation}
Then, the entanglement entropy (\ref{eq:SreducedR}) simply reads
\begin{equation}
S(t) = \sum_j e(  \gamma_{j} ),\; e(x) \equiv -x\ln x - (1\!-\!x) \ln(1\!-\!x).
\label{eq:S}
\end{equation}
This procedure (see \cite{calabreseJSM4} for details)
results in an efficient numerical method which essentially only requires
diagonalization of $2L$ dimensional matrix $\Gamma$ for the solution of a quantum problem
on $2^{4L}$ dimensional Hilbert space.

The time-dependent a-fermi operators $\ww_m(t)$ are obtained 
from {\em linear} Heisenberg type equations ${\dot \ww_m} = -\ii [\ww_m, \mathcal{H}]$,
namely,
${\dot \ww_{2j}}=2(\ww_{2j+1}-h\ww_{2j-1})$ and
${\dot \ww_{2j-1}}=2(-\ww_{2j-2}+h\ww_{2j})$.
The solution of such Heisenberg equations, written as 
$ {\dot \ww_m} = -\ii\sum_n G_{mn} \ww_n$,
is obtained by diagonalizing a $2L\times 2L$ matrix 
$G = V \cdot \Lambda \cdot V^{\dagger}$ 
which yields
\begin{equation}
	\ww_m(t) = \sum_n \Big( \sum_k V_{mk} \ee^{-\ii t \Lambda_k} V_{nk}^* \Big) \ww_n.
\end{equation}

However, in the `critical case' $h=1$, 
the time-evolution of $\ww_m(t)$
can be solved exactly and some analytical
solutions can be given.
Namely in such a case the sets of Heisenberg eqs. 
are identical and are solved via
discrete sine-transform 
with 
$V_{mk}=\ii^{m}\sqrt{\frac{2}{4L+1}}\sin\big[\frac{(m\!+\!2L)k\pi}{4L+1}\big]$,
\begin{equation}
	\ww_m(t) = \sum_n \Big[ 
	\sum_{k=0}^{4L} V_{mk} \ee^{\ii 4 \cos(\frac{k\pi}{4L+1}) t} V_{nk}^{*}
	\Big] \ww_n.
	\label{eq:discretefoursolution}
\end{equation}

In the following we shall be interested in the results in the 
{\em thermodynamic limit} (TL), $L\to\infty$.
The infinite sum over $k$ in (\ref{eq:discretefoursolution}) 
is transformed onto an integral which yields
$\ww_{m}(t) = \sum_{n\in\ZZ}\Phi_{nm}(4t) {\hat w_n}$ in terms of Bessel functions
$\Phi_{ab}(x) \equiv J_{a-b}(x)$.
The correlation matrix elements are therefore 
(using ${\hat n_b} = {\ww_b}^\dagger {\ww_b}$)
\begin{equation}
	\Gamma_{mn}(t) = \sum_{b\in \ZZ}\Phi_{bm}(4t)\Phi_{bn}(4t)
	\xbra{A}{\hat n_b}\xket{A}.
\end{equation}
We also assume that the initial operator $A$ is {\em local}, i.e. a product of {\em finite} number of Pauli matrices $\sigma^\alpha_j$. This implies that: (i) either
$A$ has a {\em finite} index, i.e. $\xave{\nn_b}\equiv \xbra{A}\hat{n}_b\xket{A} = 0$, for $|b| > b_0$ for some $b_0 \in \ZZ^+$, or (ii) $A$ has an infinite index and $\xave{\nn_b}=1$ for $b < -b_0$ and 
$\xave{\nn_b}=0$ for $b > b_0$ (such as e.g. $A=\sigma_1^{\rm x}$).
Then as $L\to \infty$, an arbitrary large fixed finite piece of correlation 
matrix can be asymptotically written as
\begin{equation}
	\Gamma_{mn}(t) = \sum_{b\in \ZZ}
	J_{b-m}(4t)J_{b-n}(4t)
	\xave{\hat n_b}.
	\label{eq:approximationgamma}
\end{equation}
Note that $\Gamma_{mn}$ has effectively finite rank $\sim x=4t$, namely 
$\Gamma_{m,n} \sim \delta _{mn}$, for $-m,-n > x$.
For brevity we shall be omitting the argument of Bessel functions 
always equal to $x=4t$.

\subsection{Initial operators of finite index}

First, we focus on the case (i) of finite index initial operators $A$.
It was conjectured in \cite{prosenPRE75} that in such cases the entanglement entropy 
in thermodynamic limit saturates in time.
Using the a-fermion algebra we are now able to calculate the exact
saturation value since RHS in~(\ref{eq:approximationgamma}) 
is a finite sum.
Consider $\Gamma_{mn}$ as a real matrix over $\RR^{\infty}$ with canonical
basis $\{\ket{m},m\in\ZZ^-\}$ and write a set of {\em non-orthogonal} vectors
$\{\ket{\psi_\alpha}\}$, namely $\braket{m}{\psi_\alpha} = (-1)^{\alpha} J_{\alpha-m}(4t) = \braket{\psi_\alpha}{m}$.
Let us write initial operator of finite index ($K$) as 
$A =  O_{j_1}\cdots O_{j_K}$ where $O_{2j-1}\equiv X_j$ and $O_{2j}\equiv Y_j$.
Then we have 
$\Gamma_{mn} = J_{j_1-m}J_{j_1-n} + \cdots + J_{j_K-m}J_{j_K-n}$, or
\begin{equation}
	\Gamma_{mn}(t) = 
\braket{m}{\psi_{j_1}}\braket{\psi_{j_1}}{n} + 
\cdots + 
\braket{m}{\psi_{j_K}}\braket{\psi_{j_K}}{n}.
\label{eq:operatorfinitegamma}
\end{equation}
This means that the rank of $\Gamma_{mn}$ is bounded by $K$, in fact it is $K$, and
its non-trivial eigenspaces are spanned by $\{\ket{\psi_{j_k}},1\le k \le K\}$.
Let $\{\ket{\phi_k},1\le k \le K\}$ be an orthonormalized set obtained from  
$\{\ket{\psi_{j_k}},1\le k \le K\}$ by a standard Gramm-Schmidt procedure, for which the
only input is the set of scalar products
$\braket{\psi_\alpha}{\psi_\beta} = 
\sum_{k\in \ZZ^-} J_{k-\alpha}J_{k-\beta}$
which can be in TL evaluated analytically for {\em any} $t$ in terms of finite sums and
approach the following long time asymptotics
$\braket{\psi_\alpha}{\psi_\beta}\vert_{t=\infty} = (1/2)\delta_{\alpha\beta} - \sin[\pi(\alpha-\beta)/2]/[\pi(\alpha-\beta)]$.
The non-vanishing part of the spectrum $\{\gamma_j\}$ of the correlation matrix
(\ref{eq:operatorfinitegamma}) entering eq. (\ref{eq:S})  
is thus given by the eigenvalues of the following $K\times K$ matrix
\begin{equation}
	\tilde{\Gamma}_{kl} = 
\braket{\phi_k}{\psi_{j_1}}\braket{\psi_{j_1}}{\phi_{l}} + 
\cdots + 
\braket{\phi_k}{\psi_{j_K}}\braket{\psi_{j_K}}{\phi_{l}}.
\label{eq:gammatilde}
\end{equation}
Thus, eq. (\ref{eq:gammatilde}) is our main result for the case of finite index initial operators.
\begin{figure}
	\centering
	\includegraphics[width=\columnwidth]{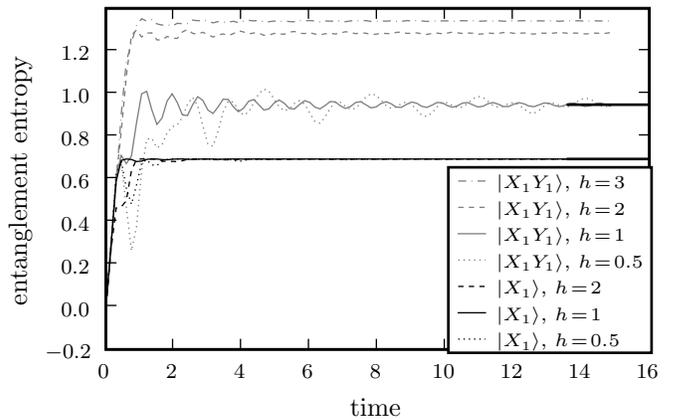}
	\caption{Entanglement entropy for finite-index operators $\xket{X_1}$ (black)
	and $\xket{X_1 Y_1}=\ii \xket{\sigma^{\rm z}_1}$ (gray) compared to theoretic saturation value 
	for $t\to\infty$ and $h=1$ (thick lines). Three different values of magnetic field
	are considered:  $h=1$ (solid curve), $h=0.5$ (dotted), $h=2$ (dashed), $h=3$ (dash-dotted).
	We set $2L=200$, such that no finite size effect were noticable.
 	}
	\label{fig:figA}
\end{figure}
For illustration, let us calculate the asymptotic value $S(t\to\infty)$ 
for the simplest two cases: (a) $A=O_j$, e.g. 
$A=\cdots \sigma^{\rm z}_{-2}\sigma^{\rm z}_{-1}\sigma^{\rm x}_0$, 
and (b) $A= O_j O_{j+1}$, e.g. $A=\sigma^{\rm z}_1$.
In case (a), $K=1$, the result is $\gamma_1 = \braket{\psi_j}{\psi_j} $ with 
$\gamma_1\vert_{t=\infty} = \frac{1}{2}$
which gives the entanglement entropy of $S(\infty) = \ln 2$.
In case (b), $K=2$, we have 
$\gamma_{1,2}=\frac{1}{2}[\braket{\psi_j}{\psi_j}+\braket{\psi_{j+1}}{\psi_{j+1}} \pm \sqrt{ (\braket{\psi_j}{\psi_j}+\braket{\psi_{j+1}}{\psi_{j+1}})^2+4\braket{\psi_{j}}{\psi_{j+1}}^2}]
$ with $\gamma_{1,2}\vert_{t=\infty} =    \frac{1}{2}\pm \frac{1}{\pi}$
and $S(\infty) = 2 \ln(\gamma_1^{-\gamma_1}\gamma_2^{-\gamma_2})$.
Both results agree excellently with numerical solutions of Heisenberg eqs.
for $\ww_j(t)$ shown in  Fig.~\ref{fig:figA}, for the case $h=1$, whereas saturation is observed 
for any $h$. 

\subsection{Initial operators of infinite index}
  
Second, we consider the case (ii) of infinite index initial operator $A$.
In thermodynamic limit, local spin operators such as $\sigma_1^{\rm x}$ are products of infinite number of
Majorana operators $X_n$, $Y_n$, in particular 
$\xket{\sigma_1^x}=\xket{\cdots X_{-1} Y_{-1} X_0 Y_0 X_1}$ and
previous disciussion does not apply.
As conjectured in~\cite{prosenPRE75} time complexity
for such initial operators only grows polynomially in time
which corresponds to logarithmic growth of the entanglement entropy.
\begin{figure}
	\centering
	\includegraphics[width=\columnwidth]{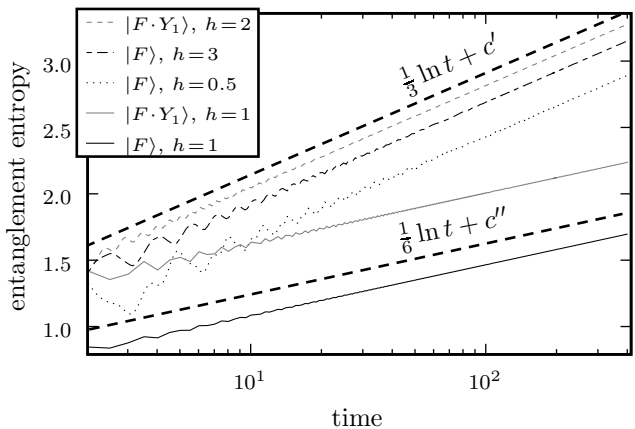}
	\caption{Entanglement entropy for infinite-index operators
	$\xket{F}$ (black) and $\xket{F Y_1}=\xket{\sigma_1^{\rm y}}$ (gray) 
	and different magnetic fields (same styling as in Fig.~\ref{fig:figA})
	as they appear in the legend from top to bottom. 
	Thick dashed lines correspond to $(1/3)\ln t$ and $(1/6)\ln t$.
	}
	\label{fig:figB}
\end{figure}
Let us define an infinite index operator $F = \cdots X_{-1}Y_{-1}X_0 Y_0$ which corresponds
to a half-filled Fermi sea $\xket{F}$ of a-fermions. Any operator of interest here can be written
as $A = F B$ where $B$ is a finite-index operator, again $\xket{A}$ can be intepreted as
a Fermi sea with a finite number of particle and hole excitations.
Figure~\ref{fig:figB} shows results for $S(t)$ (\ref{eq:S}) based on
numerical solution of Heisenberg equations for $\ww_j(t)$ for $2L$ up to $800$
such that no finite size effects are noticable in the figure. For any initial operator of the form
$A = F B $, we observe a clean asymptotic logarithmic growth
\begin{equation}
S(t) \asymp c \ln t + c', 
\textrm{  where }
c = \left\{ \begin{array}{ll}
1/6 & \textrm{ if } |h| = 1; \\
1/3 & \textrm{ if } |h| \neq 1,
\end{array}
\right.
\label{eq:Sc}
\end{equation}
and $c'$ is a constant which, for given $h$, only depends on the choice of finite index operator $B$.
Note an intriguing similarity with the size scaling of entanglement entropy of the ground 
state of (\ref{eq:H}) \cite{latorre,calabreseJSM2004}.
Analytical explanation of this interesting phenomenon may be non-trivial, so in the
following we limit ourselves to the case of critical field $h=1$ and consider only the
simplest initial operator of infinite index, namely $A=F$ where the problem can be connected
to the theory of block Toeplitz determinants.
In order to avoid negative indices, we re-define the correlation matrix as $\Gamma'_{mn} \equiv
\Gamma_{-m,-n}$, so from eq. (\ref{eq:approximationgamma}) follows
\begin{equation}
\Gamma'_{mn} = (-1)^{m+n}\sum_{b=0}^\infty J_{b-m}J_{b-n},
\quad
m,n=0,1,\ldots
\label{eq:gammaZZ}
\end{equation}
It should be noted that correlation matrix can be factorized 
$\Gamma'_{mn} = \sum_{l=0}^\infty \Psi_{m l} \Psi_{l n}$ as a square of a matrix
$\Psi_{mn} = (-1)^n J_{m-n}$.
Note that $\Psi_{mn}$ is a real symmetric infinite block Toeplitz matrix
\begin{equation}
\Psi = 
\begin{pmatrix} \Pi_0 & \Pi_1 & \ddots \\
                \Pi_{-1} & \Pi_0 & \ddots \\
                \ddots & \ddots & \ddots 
                \end{pmatrix},\quad
\Pi_l \equiv \begin{pmatrix} 
J_{2l} & J_{2l+1} \\
-J_{2l-1} & -J_{2l} 
            \end{pmatrix}.                       
\end{equation}
Following Ref. \cite{jinJSP116} we express the time-dependent entanglement entropy (\ref{eq:S})
in terms of a formula involving Block Toeplitz determinant
\begin{equation}
	S = \frac{1}{2\pi \ii} \int_{\Xi} e(\lambda^2) \big[ 
	\frac{\dd}{\dd \lambda}\ln \det(\lambda\one - \Psi)
	\big] \dd \lambda
	\label{eq:residua}
\end{equation}
where $\Xi$ is a closed curve in complex plane enclosing unit disk, avoiding 
point $1$ and interval $[-1,0]$.
Note that eigenvalues of infinite dimensional matrix $\Psi$ come in pairs $\lambda,-\lambda$ with
accumulation points $\pm 1$.
For any $\epsilon > 0$ there is only a finite number,
 $N_\epsilon(t) \sim t$, of eigenvalues of $\Psi$ which are not in $\epsilon$ vicinity of $\pm 1$.
However,  we find numerically that 
most of these eigenvalues cluster around $0$, and only $\sim \ln t$ of them
lie outside $\epsilon$ vicinity of $0$ which are the only eigenvalues contributing to entanglement
entropy result (\ref{eq:Sc}).

At present state of the theory of block Toeplitz determinants - in connection to the
theory of integrable Fredholm operators and the Riemann-Hilbert problem \cite{deiftAMST} 
- the formula (\ref{eq:residua}) can be explicitly evaluated (see e.g. Ref.\cite{itsJPA38})
provided the {\em matrix symbol}
$\Phi(z) = \lambda \one - \sum_{k\in\ZZ} \Pi_k z^k$, 
admits explicit Wiener-Hopf factorizatons
$\Phi(z) = U^+(z) U^-(z)=V^-(z) V^+(z)$ where the matrix functions $U^{\pm}(z),V^{\pm}(z)$ 
are analytic inside($+$)/outsize($-$) the unit circle.
Even though the matrix symbol has an appealing explicit form 
\begin{equation}
	\Phi(z) = \left( \begin{array}{cc}
		\lambda - f \! \bar{f} + g \bar{g} & f \bar{g}/z - g \bar{f} \\
		z g \bar{f} - f \bar{g} & \lambda +  f \! \bar{f} - g \bar{g} 
		 \end{array} \right)
\end{equation}
where $f=f(z),\bar{f}=f(z^{-1}),g=g(z),\bar{g}=g(z^{-1})$ and
 $f(z) \equiv {\rm cosh}(2t \sqrt{z})$ and $g(z) \equiv {\rm sinh}(2t \sqrt{z})/\sqrt{z}$ 
 are {\em entire} analytic functions, its Wiener-Hopf 
 factorizaton is at present unknown and poses a challenging problem.
  
\section{Conclusions}
\label{sect:conclusions}

We have studied complexity of time evolution of initial local operators under dynamics given
by the transverse Ising chain. Such complexity can be characterized
in terms of entanglement entropy of operators treated as elements of a product Hilbert space
corresponding to a bi-partition of a chain and is directly related to time efficiency of simulation
methods such as tDMRG. Note that operator space entanglement entropy, of
say a density operator, is {\em not}
simply related to a traditional notion of entanglement of the corresponding mixed state.
For example, consider a macroscopic convex combination of $~2^L$ product states. 
This corresponds to a {\em non-entangled} mixed state but has
a macroscopic operator space entanglement entropy $\sim L$ 
and hence it is difficult to simulate classically.
Thus it seems that the traditional concept of 
{\em state entanglement} is not sufficient to
characterize classical complexity of quantum operators.
In this paper we have shown, in parts analytically and numerically, 
that operator space entanglement entropy of transverse Ising model grows 
at most logarithmically for initial operators which are local products of Pauli matrices. 
This result has to be contrasted with a a linear growth of entanglement
entropy for time evolution of  pure states \cite{calabreseJSM4}. 
Explanation of deeper physical reasons for this dramatic effect is needed.

    Stimulating discussions with J. Eisert and M. \v Znidari\v c and 
    support by the grants P1-0044 and J1-7347 of Slovenian Research Agency
    are gratefully acknowledged.

\end{document}